\documentstyle[12pt]{article}
\def\be{\begin{equation}}
\def\ee{\end{equation}}
\def\la{\label}

\begin{document}

\title{Gaugino condensation in 4-D superstring models}
\author{A. de la Macorra and G.~G.~Ross\thanks{SERC Senior
Fellow}}
\date{{\small Department of Physics, University of Oxford,\\
  1 Keble Rd, Oxford OX1 3NP}}
\maketitle

\begin{abstract}
\noindent
We study the possibility that supersymmetry is broken via a
gaugino condensate in four dimensional string theories. We derive
an effective low-energy theory describing the Goldstone mode
associated with the R-symmetry breaking driven by gaugino
condensation and show that this gives a description of gaugino
condensation equivalent, at tree level, to other  effective
Lagrangian approaches based on the symmetries of the theory. A
study of the mass gap equation
shows that non-perturbative effects favour the appearance of a
gaugino condensate. By minimizing the effective potential
describing both the condensate and the moduli determining the
gauge coupling constant and radius of compactification we show
that large hierarchy may occur between the gravitino mass and the
Planck mass with the existence of only one gaugino condensate.
Associated with this is a determination of the fine structure
constant and the compactification scale.
\end{abstract}

\section{Introduction}

There has recently been considerable attention focused on the
study of supersymmetry (SUSY) breaking in the effective
Lagrangian  obtained from superstring compactification \cite{x0}.
If one is to avoid the inclusion of an explicit mass scale
associated with this breaking then it must proceed through
nonperturbative effects\cite{Dine/Seiberg}. Perhaps the most
promising origin for such breaking is via gaugino condensate
\cite{x1,2,3} in the hidden sector for it can easily lead to a
large hierarchy between the Planck and gravitino mass. This is
so because the scale of gaugino condensation is expected to
correspond to the scale at which the gauge coupling constant
becomes strong. If the coupling at the
compactification scale is small then, using the renormalization
group equation, the scale at which the coupling becomes large is
exponentially suppressed relative to the initial scale.

In order to study the SUSY breaking which results from gaugino
condensation it is necessary to determine the effective low-
energy theory below the scale $\Lambda_{gut}$ of string
compactification. Most analyses start by introducing
 a ``truncated superpotential'' $W = <S\lambda\lambda>$, where
S is the modulus superfield determining the gauge coupling,
$g^2=\frac{1}{Re S}$, at the compactification scale, $\lambda$
the gaugino field and the brackets denote the vacuum expectation
value (v.e.v.).
The form of
$<S\lambda\lambda>$ has been  determined by dimensional analysis,
by instanton calculations \cite{5} and by imposing an R-symmetry to
the superpotential \cite{3,4,5}.  These all require
$<S\lambda\lambda>$ to have the form $c\,exp(-3\,S/2\,b_{0})$
where $c$ is proportional to $\Lambda^{3}_{gut}$ and $b_{0}$ is
the coefficient of the $g^{2}/4\pi$ term in the beta-function
associated with the hidden sector gauge group.
Another approach to parameterize the gaugino condensate  is via
the effective Lagrangian \cite{6}.
In this approach the gaugino bilinear is assigned to a chiral
superfield and its superpotential is obtained by demanding that
it gives the correct terms to cancel the trace, axial and
superconformal anomalies. The functional dependence of the
gaugino condensate, Y , on the modulus fields S and T is given
by minimizing the scalar potential.

Although these approaches give a general parameterization of the
gaugino condensate they do not address the dynamical question why
a condensate is energetically favoured. In practice this is an
important consideration for the contribution to the vacuum energy
from gaugino binding effects can play a significant role in
determining the structure of the potential and the SUSY breaking
effects in the visible sector. In order to study such effects it
is necessary to evaluate the nonperturbative effects giving rise
to gaugino condensation. The complete solution is clearly beyond
our present-day technology so we are forced to employ
approximation methods. Here we will apply Nambu-Jona-Lasinio
(N-J-L) techniques \cite{7} to obtain non-perturbative information
about the gaugino binding.

We start by constructing the effective Lagrangian describing the
Goldstone mode that results when a gaugino condensate forms due
to the spontaneous symmetry breaking of R-symmetry. The form of
the effective Lagrangian describing the Goldstone mode is
strongly constrained by the requirement of N=1 local SUSY
\cite{9} and we show that this leads to a prediction of the
dependence of the gaugino condensate on the modulus S and T
consistent with the previous approaches mentioned above. However
our approach shows that the resultant tree-level form with an
effective four-fermion coupling should be radiatively corrected.
In non-susy models it is known that, for sufficiently strong
coupling, such a 4-Fermi interaction drives a fermion
condensate dynamically and breaks the chiral symmetry
spontaneously. By contrast, in the global SUSY version \cite{8}
the formation of bound states is not dynamically preferred, due
to SUSY.  In the local SUSY model obtained from 4-D superstring
of interest here, a (non-perturbative) calculation of the
radiative corrections via the Schwinger-Dyson equation shows that
dynamical symmetry breaking is energetically favoured.  The
effective potential from the S and T fields is shown to be
bounded from below in a simple Orbifold model once the
constraints of duality invariance \cite{10,18,d} are satisfied
(The inclusion of this symmetry is relevant for fixing the vacuum
expectation value (v.e.v.) of the modulus T that parameterizes
the compactified dimensions). By minimizing the full effective
potential for the moduli and gaugino we show that a large mass
hierarchy may develop, with a reasonable prediction for the
compactification scale and the value of the fine structure
constant.

The outline of the paper is as follows. In Section 2 we introduce
the structure of the effective Lagrangian in the hidden sector
of a four dimensional superstring theory and the consider the
expected form of the gaugino condensate. By way of motivation we
discuss in Section 3 the N-J-L analysis of fermion condensation
in the non-supersymmetric case. In Section 4 we derive an
effective Lagrangian which describes the Goldstone mode of broken
R-symmetry and show that it leads to a parameterisation of the
gaugino condensate equivalent, at tree level, to previous
approaches. We conclude this section with a calculation of the
1 loop corrections. In Section 5 we discuss the N-J-L analysis
in the locally supersymmetric case and show that, taking the
constraints of duality into account, there is a stable minimum
at which the scale of SUSY breaking is determined. Section 6
presents our conclusions.

\section{The effective Lagrangian in D=4 superstrings}

The effective D=4 superstring inspired model \cite{x0} is a N=1
supergravity (sugra) \cite{9} with at least two gauge-singlet
moduli S and T as well as an unspecified number of gauge chiral
matter superfields. The N=1 sugra is specified by
two functions, the Kahler potential, G, and the gauge kinetic
function, f. In the simplest case a single (1,1) modulus
superfield T
determines the radius of the compactified space.
The Kahler potential is given by (here and henceforth we set
$M_{Planck}=1$)
\be
G=K-ln(\frac{1}{4}|W|^{2})
\la{e1}\ee
where $K=ln(S+\bar{S})+3ln(T+\bar{T}-2\mid\varphi_{i}\mid^{2})$,
$\varphi_{i}$ are the (untwisted) chiral superfields and $W$ is
the superpotential that depends on the chiral superfields.  The
gauge coupling constant is given in terms of the real part of the
gauge kinetic function,  $g^{-2}= Ref$, and at the
compactification scale it is just $g^{-2} = ReS$.

As we will discuss in Section \ref{sec-mod} the form of the
superpotential should be restricted by the modular invariance of
the underlying string theory. In order to make contact with
previous results we first discuss in Section \ref{sec-efflag} the
non-modular invariant form (equivalent to dropping the
contributions of the Kaluza Klein and winding modes) and
subsequently determine the full modular invariant form in
Section \ref{sec-mod}.

It is believed that a gaugino condensation will form at a scale
where the gauge coupling constant becomes strong. This scale is
determined by the renormalization group equation and at the one
loop level it is \be
\Lambda_{c}^{2}=\Lambda_{gut}^{2}e^{-ReS/b_{0}}
\la{e2}\ee
with
\be
\Lambda_{gut}^{2}=\frac{1}{<ReS ReT>}
\la{e3}\ee
the compactification scale. From dimensional analysis or
instanton calculations we know that the gaugino condensate, if
it occurs, should be proportional to $\Lambda_{c}^{3}$. Note that
gaugino condensation, if it forms below the compactification
scale, is largely a field theory phenomenon coming from the
strong Yang-Mills gauge forces in the hidden sector far below the
scale of string excitations. The difference from usual Yang-Mills
theory lies in the fact that the gauge coupling is itself a
dynamical variable ($g^{-2}\propto Re S$). The value of the vev
for the modulus S determines the value of the gauge coupling at
the compactification scale and hence the magnitude of the
hierarchy.

 Here we wish to discuss whether non-perturbative effects favour
a  gaugino condensate and to determine the effective potential
for the S field. The approach we employ is first to construct the
effective Lagrangian describing the light degrees of freedom
below the scale at which the gauge coupling becomes strong.
If the effect of this interaction is to generate a gaugino
condensate, the R-symmetry under which the gauginos transform
will be spontaneously broken. The effective Lagrangian below
$\Lambda_{c}$ will then describe the Goldstone modes associated
with this spontaneous symmetry breaking and the stability or
otherwise of the system to such breaking can be discussed in
terms of these fields. This  approach closely parallels the
procedure adopted in the N-J-L model and by way of motivation we
first discuss how the analysis proceeds in the non-supersymmetric
case.

\section{The non-supersymmetric N-J-L model}
\label{sec-njl}

The non-SUSY N-J-L model starts with a four-fermion interaction
described by the Lagrangian given by \cite{7}
\be
L=i\bar{\psi}\gamma^{\mu}\partial_{\mu}\psi+\frac{1}{4}g^{2}((
\bar{\psi}\psi)^{2}-(\bar{\psi}\gamma_{5}\psi)^{2})
\la{e4}
\ee
or in two component notation \footnote{We define
$\bar{\psi}_{R,L}=\frac{1}{2}(1\pm\gamma_{5})\psi,
\bar{\psi}_{R}\equiv(\psi_{R})^{\dagger}\gamma_{0}$}
\be
L=i(\bar{\psi}_{L}\sigma^{\mu}\partial_{\mu}\psi_{L}
+\bar{\psi}_{R}\sigma^{\mu}\partial_{\mu}\psi_{R})
+g^{2}\bar{\psi}_{L}\psi_{R}\bar{\psi}_{R}\psi_{L}.
\la{e5}
\ee
Here $g^{2}$ is a dimensional coupling, $g^2=h^{2}/\Lambda^{2}$,
where h is a dimensionless  and $\Lambda$ is the mass scale at
which the new physics generating the four-fermion interaction
appears. The theory has a $U(1)_{L}\otimes U(1)_{R}$ chiral
symmetry of independent phase rotations of the left and right
handed fermion components. Eq.(\ref{e5}) can be rewritten in
terms of an auxiliary scalar field $\phi$, \be
L=i(\bar{\psi}_{L}\sigma^{\mu}\partial_{\mu}\psi_{L}+\bar{\psi
}_{R}\sigma^{\mu}\partial_{\mu}\psi_{R})-
|\phi|^{2}+g\phi^{\ast}\bar{\psi}_{R}\psi_{L}+g\phi\bar{\psi}_
{L}\psi_{R} \la{e6}\ee
and by its classical equation of motion $\phi$ is identified with
$g\bar{\psi}_{R}\psi_{L}$. Eliminating $\phi$ just gives
eq.(\ref{e5}).

The tree level potential
\be
V_{0}=|\phi|^{2}
\la{e7}\ee
is semipositive definite and the minimum is at $\phi=0$, i.e. no
condensation state. The one-loop corrections are properly taken
into account by the Coleman- Weinberg result \cite{13}
\be
V_{1}=-\frac{1}{8\pi^{2}} \int d^{2} p\, p^{2}
ln(p^{2}+m^{2}_{F}). \la{e8}\ee
Integrating eq.(\ref{e8}) out using a momentum-space cutoff,
because the interaction is non-renormalizable, one obtains
\be
V_{1}=-\frac{\Lambda^{4}}{16\pi^{2}}(x+x^{2}
ln(\frac{x}{1+x})+ln(1+x)) \la{e9}\ee
with
\be
x=\frac{m^{2}_{F}}{\Lambda^{2}}=
\frac{4g^{2}|\phi|^{2}}{\Lambda^{2}}.
\la{e10}
\ee
The scalar potential is then the sum of $V_{0}$ and $V_{1}$ and
it is given by \be
V=\frac{\Lambda^{4}}{16\pi^{2}}(\frac{2}{\alpha} x-x-x^{2}
ln(\frac{x}{1+x})-ln(1+x)) \la{e11}\ee
with
\be
\alpha=\frac{g^{2} \Lambda^{2}}{8\pi^{2}}.
\la{e12}\ee
{}From eq.(\ref{e11}) it is easy to see that the extremum condition
is \be
\frac{\partial V}{\partial
\phi^{\ast}}=\phi\alpha(\frac{1}{\alpha}-1-xln(\frac{x}{1+x}))=0.
\la{e13}\ee
Provided V has a negative slope at the origin  eq.(\ref{e13})
admits a non-trivial solution which is dynamically preferred.
This is possible only for
\be
\alpha=\frac{g^{2} \Lambda^{2}}{8\pi^{2}}>1,
\la{e14}\ee
i.e. a strong coupling constant. In this case eq.(\ref{e13})
gives \be
1=\frac{g^{2}\Lambda^{2}}{8\pi^{2}}(1+\frac{m^{2}_{F}}{\Lambda
^{2}}\,ln(\frac{m^{2}_{F}/\Lambda^{2}}{1+m^{2}_{F}/\Lambda^{2}}))
\la{e15}\ee
which is the familiar mass gap equation that could have
been derived from the interaction of eq.(\ref{e4}). Note that
this solution is necessarily non-perturbative since it equates
tree level and one loop contributions. It is straightorward to
show that it amounts to a non-perturbative summation of fermion
bubble graphs, which are dominant in the large $N_{c}$ limit
where $N_{c}$ is the number of colours.

The solution corresponds to the case  where $\phi$ acquires a
v.e.v. and the $U(1)_{L}\otimes U(1)_{R}$ chiral symmetry of
the Lagrangian eq.(\ref{e4}) is broken to $U(1)_{L+R}$ . In this
case the associated Goldstone mode may be identified with the
field $\phi$ which, through quantum effects, acquires a kinetic
term $L_{k}$ and becomes a propagating field. It may be seen that
$L+L_{k}$ is the effective Lagrangian describing the light
(Goldstone) degrees of freedom, appropriate below $\Lambda$,
together with the fermion field. The effective potential
calculated using this Lagrangian just reproduces the results
using the more familiar Schwinger-Dyson equation following from
the original Lagrangian eq.(\ref{e4}).  Thus the effective
Lagrangian describing the
would-be Goldstone mode provides a convenient way of implementing
the N-J-L scheme for summing the leading terms in the large
$N_{c}$ limit.

\section{An effective Lagrangian description of gaugino
condensation}
\label{sec-efflag}
We turn now to the superstring inspired N=1 sugra model. As in
the case of the N-J-L model we look for a formulation that
parameterizes the gaugino bilinear by a classically non-
propagating auxiliary field which at the quantum level becomes
the Goldstone mode associated with the breaking of a continuous
symmetry. In this case, however, the formation of the original
theory in terms of an auxiliary field must be made consistent
with local SUSY.

\subsection{The broken R-symmetry Goldstone mode}
\label{sec-gobo}
For gauginos, in the absence of superpotential terms, there is
an R symmetry which is spontaneously broken if a gaugino
condensate forms leading to a Goldstone mode. In this case the
auxiliary field $\phi$ describing this would-be mode must be
embedded in a chiral superfield $\Phi$ which is coupled in a
supersymmetric  way.

For a superpotential $W$ and gauge kinetic function $f$ depending
on an auxiliary chiral superfield $\Phi=(\phi,\chi,h)$, where
$\phi,\chi,h$ are the scalar, fermion and auxiliary field
components, the terms in the N=1 sugra lagrangian involving these
fields are (suppressing all gauge indices) \cite{9} :
\[
L_{aux}=\frac{1}{2} \frac{\partial W}{\partial  \phi} h -
\frac{e^{K/2}}{4} \frac{\partial f}{\partial \phi}
\bar{\lambda}_{R} \lambda_{L} h + (
\frac{1}{2}\bar{\psi}_{L}\cdot\gamma W^{i}\chi_{Li}-
\frac{1}{2}e^{K/6}W^{ij} \bar{\chi}_{Ri} \chi_{Lj}  \]
\be +
\frac{1}{4}e^{2K/3}\bar{\lambda}_{R}\lambda_{L}f^{ij}\bar{\chi
_{Ri}}\chi_{Lj}-
\frac{1}{4}e^{K/2}f^{i}\chi_{Li}\bar{\psi_{L}}\cdot\gamma\bar{
\lambda}_{R}\lambda_{L} + h.c.) \la{e16}\ee
where $\lambda_{L}$ represents the gaugino
field\footnote{Again we define
$\lambda_{R,L}=\frac{1}{2}(1\pm\gamma_{5})\lambda,
\bar{\lambda}_{R}\equiv(\lambda_{R})^{\dagger}\gamma_{0}$} (with
kinetic term
$L_{k}=\frac{i}{2} Ref\,
\bar{\lambda}\gamma^{\mu}\partial_{\mu}\lambda $), $\psi$ the
gravitino field, $\chi_{i}$ the fermion component of the chiral
matter superfield $\varphi_{i}$  and $W_{i}\equiv \frac{\partial
W}{\partial z^{i}}$ with $z^{i}$ the scalar component of
$\varphi_{i}$. The classical equation of motion for $\Phi$
yield\\
i)$\frac{\delta L}{\delta h}$
\[
\frac{1}{2}\frac{\partial W}{\partial
\phi}=\frac{e^{K/2}}{4}\frac{\partial f}{\partial \phi}\,
\bar{\lambda}_{R}\lambda_{L}
\]
ii)$\frac{\delta L}{\delta \chi}$
\[
\bar{\psi}_{L} \cdot \gamma \,(\frac{1}{2}\frac{\partial
W}{\partial \phi}-\frac{e^{K/2}}{4}\frac{\partial f}{\partial
\phi}\, \bar{\lambda}_{R}\lambda_{L})
+e^{K/6}(\frac{e^{K/2}}{4}\frac{\partial f^{i}}{\partial
\phi}\bar{\lambda}_{R}\lambda_{L}-\frac{1}{2}\frac{\partial
W^{i}}{\partial \phi})\,\bar{\chi}_{Ri}=0
\]
and\\
iii)$\frac{\delta L}{\delta \phi}$
\be
\frac{\partial V}{\partial \phi}=0
\la{e17}\ee
with $V = L - L_{k}$ and $L_{k}$ the kinetic Lagrangian.

The effective theory describing the interaction of $\Phi$ is
specified once $W$ and $f$ are given. The
effects of gauge boson-gaugino
interaction will be to generate an effective four fermion vertex.
The form factor for this vertex vanishes rapidly above
$\Lambda_{c}$ as the gauge coupling becomes weak and is
essentially constant below $\Lambda_{c}$ for the gauge sector is
confined and all masses are of the order
$\Lambda_{c}$.

If we demand that the effective theory given in terms of the
auxiliary field $\Phi$ generates this  4-Fermi interaction then
the form of the $W$ and $f$ are uniquely determined (up to a
constant), \be
W=m^{2}\phi,
\la{e18}\ee
\be
f=\xi ln(\phi/\mu)+S
\la{e19}\ee
where $m$ and $\mu$ are mass parameters  and $\xi$ is a
dimensionless constant. From the classical equation of motion
eqs.(\ref{e17}) the scalar component of the auxiliary field
$\Phi$ is given in terms of the gaugino bilinear by
\be
\phi=\frac{e^{K/2}\xi}{2m^{2}}\, \bar{\lambda}_{R} \lambda_{L},
\la{e20}\ee
while the fermion component vanishes at the classical level. The
third equation of eqs.(\ref{e17}) is an extremum condition on the
scalar potential and once the one-loop corrections are included
it is just the mass gap equation. As we will now show, this
choice of $W$ and $f$ leads to an effective four-fermion
interaction of the desired form once the auxiliary field is
eliminated by its classical equation of motion, eq.(\ref{e20}).

\subsection{Tree level potential}

For a pure gauge theory in the hidden sector the tree level
scalar potential is  \cite{9}
\be
V_{0}=-( 3e^{-G} + h_{S}h^{S}(G^{-1})^{S}_{S} +
h_{T}h^{T}(G^{-1})^{T}_{T}) \la{e21}\ee
where $h_{S}$ and $h_{T}$   are the F-terms of the chiral
superfields S and T.  In general they are given by
\be
h_{i}=e^{-G/2}G_{i}+\frac{1}{4} f_{i}
\bar{\lambda}_{R}\lambda_{L} -G^{jk}_{i}\bar{\chi}_{Rj}\chi_{Lk}-
\frac{1}{2}\bar{\chi}_{Ri} G_{j}\chi_{L}^{j}. \la{e22}\ee
SUSY will be broken if either $h_{S}$ or $h_{T}$
 develop a non-vanishing v.e.v.. From
the structure of $V_{0}$  eq.(\ref{e21}) one can see that it is
possible to have broken SUSY and zero cosmological constant
($(G^{-1})^{i}_{j}$ are negative),  unlike in global SUSY.

The fermion partner of the auxiliary field $h_{i}$ (with i=S or
T) with non-vanishing v.e.v. will give rise to the Goldstino
field. It is easily recognized as the combination of fermions
that couple to the gravitino \cite{9} \[
\bar{\psi}_{L}\cdot\gamma\,\eta_{L}
\]
with
\[
\eta_{L}=(\frac{1}{8}f^{i}\bar{\lambda}_{R}\lambda_{L}-e^{-
G/2}G^{i})\chi_{Li}+\frac{i}{2}g G^{i}T^{j}_{i}z_{j}\lambda_{L}.
\]
In local SUSY the Goldstino will be eaten by the gravitino which
will acquire  a mass
\be
m^{2}_{3/2}=\frac{1}{4} e^{-K}|W|^{2}.
\la{e23}\ee
We may now apply this formalism to the choice of W and f given
in eq.(\ref{e18}) and eq.(\ref{e19}). In this case the F-terms
for the S and T fields are \[
h_{S}=e^{-G/2}G_{S}+\frac{1}{4} f_{S}
\bar{\lambda}_{L}\lambda_{R}  \]
\be
h_{S}=e^{-G/2}\frac{(1+S_{r}/\xi)}{S_{r}},
\la{e24}\ee
\be
h_{T}=e^{-G/2}\frac{3}{T_{r}}
\la{e25}\ee
with $S_{r}=S+\bar{S}$ and $T_{r}=T+\bar{T}$.

The tree level scalar potential is then
\be
V_{0}=m^{2}_{3/2} H
\la{e26}\ee
with
\be
H = (1+\frac{S_{r}}{\xi})^{2}
\la{e27}\ee
and the gravitino mass given by
\be
m^{2}_{3/2}=\frac{m^{4}|\phi|^{2}}{4S_{r}T_{r}^{3}}.
\la{e28}\ee
In terms of the gaugino field the scalar potential is
\be
V_{0}=\frac{\xi^{2}H}{16(Ref)^{2}}\,|\bar{\lambda}_{R}'\lambda
_{L}'|^{2} \la{e29}\ee
where the factor of $(Ref)^{2}$ in the denominator in
eq.(\ref{e29}) appears because we have rescaled the gaugino
fields appearing in this equation to have canonical kinetic
terms. Thus we have shown that a choice of $W$ and $f$ in
eqs.(\ref{e18}) and (\ref{e19}) leads to a four-fermion
interaction as desired.

The form we have derived depends on the parameters $\xi$, $m$,
and $\mu$. Since (cf.eq.(\ref{e20})) $m^{2}\phi$ is proportional to
$\bar{\lambda}\lambda$ which on dimensional grounds  we expect
to have the  v.e.v. $\propto \Lambda^{3}_{c}$ (if it develops one),
we obtain using eq.(\ref{e19})
\be
m^{2}\phi=m^{2}\mu\,e^{-
ReS/\xi}\,e^{Ref/\xi}\,\sim\,\Lambda^{3}_{c}. \la{p10}\ee
It is then naturally to chose $\xi=2b_{0}/3$ and  $m=\mu=\Lambda_{gut}$
(different choices of $m,\mu$ are possible as long as the
relation of eq.(\ref{p10}) is satisfied). As
we will see this identification is consistent with previous
parameterizations of the gaugino condensate and with the one-loop
running of the gauge coupling constant.

\subsection{Connection with parameterisations of the gaugino
condensate}

{}From eqs.(\ref{e26}-\ref{e28}) we see that a v.e.v. for $\phi$,
corresponding to gaugino condensate, will break SUSY giving the
gravitino a mass. It is worth
remarking at this point that the choice $\xi=2b_{0}/3$ means
the form of the scalar potential given by eq.(\ref{e26})
 is identical to that obtained by the ``truncated'' approach
\cite{3,4,5} in which the effects of the
gaugino condensate are included ``by hand'' by including in the
superpotential the term $W=\Lambda^{3}_{gut}exp(-3S/2b_{0})$. The
form of this superpotential was (uniquely) determined by the condition
that it should transform under the R-symmetry in the appropriate way for
a gaugino bilinear. Under the R-symmetry
\be
\lambda_{L} \rightarrow e^{-i\delta} \lambda_{L}
\la{e30a}\ee
\be
S \rightarrow S - \frac{i4b_{0}}{3} \delta.
\la{e30b}\ee
The $S$ transformation eq.(\ref{e30b}) cancels the anomalous term
($\delta L=\frac{b_{0}}{3}\,\delta\,$F\~{F}, where \~{F} is the
dual
tensor of $F$) coming from the gaugino bilinear.
In the approach adopted here when a gaugino condensate forms
$\phi$ will be the Goldstone mode associated with the spontaneous
breaking of this R-symmetry. Since the construction leading to eq.(\ref{e29})
respects both the underlying R-symmetry and local SUSY it must duplicate these
results obtained in the ``truncated'' approach which rely on the
R-symmetry. Thus we may understand the origin of the highly constrained form
of eqs.(\ref{e18}) and (\ref{e19}) leading to the potential of eq.(\ref{e29})
as following from consistency with the symmetries of the system.

In fact, it is easily seen that the effective theory described
by eq.(\ref{e18}), (\ref{e19}) and $\xi=2b_{0}/3$ transforms
correctly under the R-symmetry and it is anomalous free. From
the R-transformation of the gaugino eq.({\ref{e30a}) we deduce that
\be
\phi\rightarrow\,e^{-i2\delta}\phi
\label{p25}\ee
and the gauge kinetic function transforms as
\be
f\rightarrow\,f\,-i2\,\xi\delta
\label{p26}\ee
leaving the effective Lagrangian invariant.

It is then of no surprise that by a simple reparametrization of
the auxiliary field $\phi$ we obtain
the effective
superpotential derived by imposing that the trace, axial and
superconformal anomalies cancel at the one-loop approximation.
In this
approach an effective superpotential $P_{eff}$ is given by
\cite{6,18,8jl,10jl}
\be
P_{eff}=\frac{1}{4} Y (S + \frac{2b_{0}}{3}ln(Y/\mu'^{3}))
\la{e31}\ee
where the scalar component of $Y$ is identified with the gaugino
condensate. Its functional dependence on S is obtained by
minimizing the scalar potential and it is proportional to
$\Lambda^{3}_{c}$.

As we said before,
we may simply cast a superpotential $P$ as in this form by
defining
\be
P=-m^{2}\phi + f_{\alpha \beta}{\it w^{\alpha}w^{\beta}}
\la{p1}\ee
with $f$ given in eq.(\ref{e19}) and {\it w} the gauge covariant
chiral
superfield (which has $\lambda\lambda$ as its scalar component).
Then by   using the equation of motion for
$\phi \, (\phi= \xi {\it ww}/m^{2})$ and rescaling the auxiliary
field $\phi$,
one obtains
\be
P=\frac{m^{2}e}{\xi} \phi' (S+\xi ln(\phi' / \mu^{3}),
\la{p2}\ee
which is proportional to the superpotential used in the
``effective''
superpotential approach, eq.(\ref{e31}). Clearly the effects of
the gaugino bilinear on
 the gauge coupling obtained from eq.(\ref{e19}) and (\ref{p10})
can
be interpreted as the one-loop renormalization of the gauge
coupling given by
\be
g^{-2}(\Lambda_{c})=g^{-2}(\Lambda_{gut})+
2b_{0}ln(\Lambda_{c}/\Lambda_{gut})
\ee
and since the beta-function is  related by supergravity to the
axial, trace
and superconformal anomalies [15], this enforces the
``effective''
superpotential eq.(\ref{e31}) to take precisely the same from as
in
eq.(\ref{p2}).
\subsection{The full Effective potential}

As we have seen, the effective Lagrangian expressed in terms of
the would-be Goldstone boson correctly parameterises the form of
the gaugino condensate derived by other methods. What this
connection shows is that these analyses just give the ``tree
level'' form of the effective potential describing the gaugino
condensate and that (cf. Section ~\ref{sec-njl} ) radiative
corrections must be included. Indeed the purpose of developing
this formalism was to allow us to study non-perturbative effects
in the strong hidden sector coupling using the N-J-L method. We
proceed by calculating the $\phi$ dependence of the  effective
potential, V. If, at the minimum of V, $\phi$ develops a vacuum
expectation value it will signal that a gaugino condensate is
dynamically preferred, corresponding to the breaking of
supersymmetry. To the extent that expressing the theory in terms
of the auxiliary field $\phi$ is just a re-parameterisation of
the theory our results will be exact.

 As we have seen, eliminating $\phi$ leads to a Lagrangian
involving a four-fermion interaction. Such a four-fermion
interaction may be expected to occur due to (non-perturbative)
gauge interactions involving gauge boson and gaugino exchange
with coefficient $\propto \Lambda_{c}^{-1}$. Here we choose to
parameterise the strong gauge interaction in terms of this four
fermion interaction rather than the primary gauge and gaugino
couplings. We are then able to perform the non-perturbative sum
of these interactions corresponding to the sum of all fermion
bubble graphs.  In this way we can get, albeit incomplete,
information about the dynamics of such non-perturbative effects.
We find that they can have a dramatic effect on the structure of
the effective potential allowing for a stable non-trivial minimum
for $\phi$ corresponding to a supersymmetry breaking solution
with a large mass hierarchy. This demonstrates the importance of
including the binding effects and, at the very least, should
encourage efforts to perform a more complete summation of such
effects.

 The tree level potential for $\phi$, given by eqs.(\ref{e26}-
\ref{e28}),  \be
V_{0}=\frac{m^{4}|\phi|^{2}}{4S_{r}T_{r}^{3}}\,(1+\frac{S_{r}}
{\xi})^{2} \la{e33}\ee
is clearly semipositive
definite, as in the N-J-L model,  and the
minimum is at $\phi =0$ which implies that there is no
gaugino condensate and no SUSY breaking at tree level.
This is consistent with
the observation of Casas\footnote{ Once duality symmetry is included
the scalar potential is no longer semipositive definite at tree level
but there is no solution for reasonable values of the gauge coupling constant}
et al.\cite{21} that gaugino
condensation as usually parameterised does not occur in models
with a single hidden sector gauge group factor. However we have
argued it is essential to go beyond tree level to include non-
perturbative effects in the effective potential which may allow
for a non-trivial minimum even in the simple case of a single
hidden sector gauge group. This non-perturbative sum (equivalent
to the NJL sum) is readily obtained simply by computing the one
loop correction to V. If these destabilise the potential the
resultant minimum will correspond to a cancellation of tree level
and one-loop  terms which, as noted above, is necessarily non-
perturbative in character \cite{22}.

The one-loop radiative corrections may be calculated using the
Coleman-Weinberg one-loop effective potential,
\be
V_{1}=\frac{1}{32\pi^{2}} Str \int d^{2} p\, p^{2}
ln(p^{2}+M^{2}) \la{e34}\ee
where $M^{2}$  represents the square mass matrices and $Str$ the
supertrace. $V_{1}$  can be integrated to give
\be
V_{1}=\frac{1}{64\pi^2} \Lambda^{4} Str J(x)
\la{e35}\ee
with
\be
J( x)=x+x^{2} ln(\frac{x}{1+x})+ln(1+x)
\la{e36}\ee
and
\be
x=\frac{M^{2}}{\Lambda^{2}}.
\la{e37}\ee
Since the 4-Fermi interaction is non-renormalizable  we
regularize it by introducing  a momentum space cutoff $\Lambda$
that should be identify with the condensation scale
eq.(\ref{e2}).
The supertrace str of a function $Q(M^{2})$ is defined by \be
str Q(x)=3tr{Q(M^{2}_{A})}+tr{Q(M^{2}_{S})}-2tr{Q(M^{2}_{F})}
+2Q(4m^{2}_{3/2})-4Q(m^{2}_{3/2})
\la{e38}\ee
where $M^{2}_{A,S,F}$  are the (mass)$^2$  matrices for vectors,
scalars and spin=1/2 fields. The $2Q(4m^{2}_{3/2})$ term is the
contribution of the spin=3/2 particle, the gravitino, and
$-4Q(m^{2}_{3/2})$ is due to the gauge condition
$\psi_{R}\cdot\gamma=0$.

The scalar masses for the S and T fields are
\be
m^{2}_{S}=4m^{2}_{3/2},
\la{e39}\ee
\be
m^{2}_{T}=8m^{2}_{3/2}H.
\la{e40}\ee
In calculating the scalar masses one has to take into account
that  the scalar kinetic terms are not (and
can not be) in a canonical form. The fermion masses can
be read off of the N=1 sugra Lagrangian. The relevant terms are
\be
L_{FM}= \frac{1}{2}\bar{\chi}_{Ri} B^{ij} \chi_{Lj} + h.c.
\la{e41}\ee
with
\be
B^{mn}= m_{3/2}\,(G^{m}_{i}G^{n}_{j})^{-1/2}[D^{ij} +
\frac{1}{4\xi}(4f^{ij}-4G^{ij}_{k}(G^{-1})^{k}_{l} f^{l}-
(Ref)^{-1}f^{i}f^{j})] \la{e42}\ee
and
\be
D^{ij}=G^{ij}-G^{i}G^{j}-G^{ij}_{k}(G^{-1})^{k}_{l}G^{l}.
\la{e43}\ee
Using $W$ and $f$ of eqs.(\ref{e18}) and (\ref{e19}) the fermion
masses are then the eigenvalues of
\be
(B^{nm})=m_{3/2}\left (\begin{array}{c c}\frac{2S_{r}}{\xi}-
\frac{S^2_{r}}{4\xi Ref}  & \sqrt{3}/2 \\ \sqrt{3}/2 & 1\\
\end{array} \right ). \la{e44}\ee
Finally the gaugino mass is given by
\be
m^{2}_{g}=m^{2}_{3/2} \frac{\xi^{2} H^{2}}{4 (Ref)^{2}}.
\la{e45}\ee

These are the supersymmetry breaking masses following from the
gaugino condensate. In addition we should allow for a
supersymmetric contribution to the mass of the hidden sector
states generated by the strong hidden sector forces which (in
analogy with QCD) may be expected to be confining. Of course we
are unable to determine these masses and so we proceed by
examining the various possibilities. The first possibility is
that the gaugino condensate forms at a scale above confinement
and there is a domain in which the states are correctly described
by the gauge bosons and gauginos with the only mass coming from
the gaugino condensate as calculated above (It is thought the
equivalent situation may occur in QCD with chiral symmetry
breaking occurring before confinement). In this case  we may now
compute, using eqs.(\ref{e33}-\ref{e45}), the one loop potential.
Alternatively confinement may occur at, or above, the condensate
scale. In this case the radiative corrections should be computed
using the confined spectrum of states. Lacking knowledge of this
spectrum we may still try to estimate the result by using the
average description of these states in terms of gluons and
gluinos but allowing for the confinement effects by giving them
a common (supersymmetric) mass. We will discuss both these cases
in the next section.

\section{Dynamical breaking of SUSY}
\label{sec-7}
We are now in a position to determine whether it is energetically
favourable for a gaugino condensate to form. From eqs.(\ref{e35}-
\ref{e40}) it is clear that in order to have a SUSY breaking
solution
to the gap equation
\be
\frac{\partial}{\partial \phi}(V_{0}+V_{1})=0
\la{e46}\ee
the negative contribution from the fermion loops
must dominate (note that the contribution to the one-loop scalar
potential from each individual massive state is a monotonic
function of the mass (for a fixed cutoff) being zero only for
vanishing mass). In this limit since the (supersymmetry breaking)
gaugino mass in eq.(\ref{e45}) is proportional to $g^2$ a strong
gauge coupling constant will be dynamically preferred, i.e.
$g^{-2}=Ref <<1$. As it stands the one-loop potential will go to
$-\infty$ for $Re f=0$. This is an unphysical singularity which
will be removed when non-perturbative effects are included for
it  corresponds to infinite coupling. As we discussed in Section
4 the value of $Re f$ is not a free parameter for it defines the
initial four fermion interaction used to {\it define} the strong
binding interaction in the N-J-L approach we have adopted to
study gaugino condensation (cf. eq.(\ref{e29}). Below the scale
of gaugino condensation the effective four fermion interaction
must have the form
$\frac{c^{2}}{\Lambda^{2}_{c}}(\bar{\lambda}\lambda)^{2}$,
$c=O(1)$, where the condensation scale $\Lambda_{c}$ is also the
confinement scale. This will be true provided $[Ref(\phi)]^{-1}$
in eq.(\ref{e29}) reaches a maximum ``frozen'' value $[Ref]^{-
1}=\frac{c}{\Lambda_{c}}$. As may be seen from
eqs.(\ref{e19}),(\ref{e20}) and (\ref{p10})
the residual ambiguity in $Ref$ parameterised by c corresponds
to an ambiguity in determining $m^{2}\phi$ (and hence
$\bar{\lambda}\lambda$) in terms of $\Lambda_{c}^{3}$, relatively
unimportant when considering whether a condensate will form. We
impose this physically motivated condition  as a reasonable
parametrization of the strong coupling effects which must
eliminate the unphysical divergence associated with the vanishing
of $Ref$ and which we are presently unable to calculate .

Unfortunately this still does not cure the problem of an
unbounded potential for the potential still goes to -$\infty$
along the direction in which a gaugino condensate forms with T,
the modulus setting the radius of compactification, in the limit
$T \rightarrow 0$.
Clearly this is physically
unacceptable.  As we will see this problem may be traced to an
inadequate treatment of the effective potential in the region $T
\rightarrow 0$ (the large radius limit) which can be corrected
by demanding the potential to be invariant under the space duality
symmetry. The origin of this problem is that in the $T
\rightarrow 0$  limit corresponding to the large radius (R) limit
Kaluza
Klein modes with masses $\alpha \,1/R^{2}$ cannot be neglected
in writing the low-energy effective Lagrangian. Recent work has
shown \cite{10,11,14,18,d} that the effect of these modes may be
included by constructing a  superpotential invariant under space
duality symmetry which is known to hold for compactified string
models, and so to study the $T \rightarrow 0$ limit we turn now
to the inclusion of these Kaluza-Klein effects.

\subsection{Space-Duality symmetry and the effective potential}
\label{sec-mod}
Space duality is a symmetry of the string Lagrangian
and it is found to all orders in perturbation theory
\cite{14,18,d}.  Incorporating this symmetry in the effective D=4
theory, one has to impose duality invariance on the Kahler
potential eq.(\ref{e1}) and the gauge kinetic function f.

 Under this symmetry the S field remains invariant while the T
field transforms as an element of the SL(2,Z) group,
\be
T \rightarrow \frac{\alpha T-i\beta}{i\gamma T + \delta}
\la{e47}\ee
with $\alpha,\beta,\gamma,\delta \epsilon Z$ and $\alpha \delta -
\beta \gamma =1$. The superpotential must transform as a modular
function of weight -3, \be
W(T) \rightarrow \frac{W(T)}{(i\gamma T+\delta)^{3}}.
\la{e48}\ee
The general form for the T dependence in $W(T)$ is given in terms
of $\eta(T)$ the Dedekind eta-function ($\eta(T) =q^{1/24}
\prod_{n} (1-q^{n}), q(T)=exp(-2\pi T))$, with modular weight
1/2, and $P(T)$ , a polynomial of the absolute invariant function
$j(T)$.

The compactification scale should also be redefined so that it
is modular invariant \be
\Lambda^{2}_{gut} \rightarrow \frac{1}{<ReS ReT>|\eta (T)|^{4}}
\la{e49}\ee
With these modifications the effective N=1 sugra model and the
gauge kinetic function are modular invariant. We turn now to a
consideration of how these effects change of the locally
supersymmetric N-J-L model. In place of eqs.(\ref{e18}) and
(\ref{e26})  the superpotential and the tree level scalar
potential are now given by
\be
W=m^{2} \eta^{-6}(T) \phi,
\la{e50}\ee
\be
V_{0}=m^{2}_{3/2} H
\la{e51}\ee
with
\be
H = \frac{3T^{2}_{r}}{4\pi^{2}} |\hat{G}_{2}(T)|^{2} +
(1+\frac{S_{r}}{\xi} )^{2}-3. \la{e52}\ee
$\hat{G}_{2}(T)$ is the Eisenstein modular form with modular weight
2. Since $V_{0}$ is
modular invariant we only need to consider $T \geq 1$.

The minimum of $V$ is now well defined. A gaugino condensate is
energetically favoured, with $\phi$ acquiring a value to minimize
$Ref$. As discussed above this corresponds to
$Ref=\Lambda_{c}/c\,m_{P}$ and  \be
|\phi|^{2}=\mu^{2}\, e^{-S_{r}/\xi}\, e^{\Lambda_{c}/cm_{p}}
\la{e53}\ee
for $\Lambda_{c} << c\,m_{p}$ we have
\be
\phi = \frac{\Lambda^{3}_{c}}{\Lambda^{2}_{gut}}.
\la{e54}\ee
Using this identification the scalar potential given by
\[
V_{tot}=V_{0}+V_{1}
\]
is
\be
V_{tot}=\Lambda^{4}_{c}(\frac{H
\Lambda^{2}_{c}}{4S_{r}T^{3}_{r}|\eta|^{12}}-\frac{n_{g}}{32\pi^{2}}
J(\frac{m^{2}_{g}}{\Lambda^{2}_{c}})) + V'_{1}
\la{e55}\ee
with $J$ given in eq.(\ref{e36}), $V'_{1}$ the one-loop potential for the
$S$ and $T$ chiral superfields and $n_{g}$ the dimension of the
hidden sector gauge group.

\subsection{Determination of the Supersymmetry Breaking scale}

We have now constructed the modular invariant potential which
determine the values for the S and T fields and hence the
hierarchy relating $m_{3/2}$ to $m_{P}$ (the Planck mass). We
start with the first case discussed above in which confinement
masses are neglected. The scalar potential eq.(\ref{e55}) can be
minimized numerically. At the minimum the modulus $S$ is \be
S_{r}\simeq\frac{8\pi}{c}\sqrt{\frac{1}{n_{g}}}
\la{e56}\ee
and the gravitino (mass)$^{2}$  is
\be
m^{2}_{3/2}\simeq\frac{4\sqrt{2}}{c^{3}}(\frac{16}{3})^{3}\,
\frac{b_{0}^{9/2}}{S_{r}^{17/2}}\,e^{-3S_{r}/4b_{0}}
\la{e57}
\ee
where c is the proportionality constant in
$Ref=\frac{\Lambda_{c}}{c\,m_{p}}$  (we will take it to be one
(i.e. $c=1$)).
As seen in eq.(\ref{e56}) and (\ref{e57}) the value of the
gravitino mass is highly dependent on the one-loop beta function
$b_{0}$ and $n_{g}$. Nevertheless it is very interesting that for
certain values of $b_{0}$ and $n_{g}$  one can obtain a
phenomenologically acceptable solution with the existence of only
one gaugino condensate. Previous attempts to obtain realistic
values for the gravitino mass \cite{19,20,21} needed at least two
gaugino condensates with different gauge groups and either an
intermediate scale and/or the presence of matter fields with non-
vanishing v.e.v. As we have emphasised the difference occurs
because these analyses did not include the  radiative corrections
to the effective potential corresponding to the strong gaugino
binding effects.

We may  demonstrate that a large hierarchy may occur with
reasonable choices for the hidden sector multiplet content by
means of a simple example. We choose the gauge group SU(5) in the
hidden sector together with 6 hidden sector fermion fields in the
fundamental representation ($b_{0}=(3N-
\frac{n_{f}}{2})/16\pi^{2}$ for an SU(N) group with $n_{f}$
fermion fields in the fundamental representation). In this case
the values of the different fields, condensation scale and
gravitino mass are
\[ReS=S_{r}/2=2.65,\]
\[ReT=T_{r}/2=8.25,\]
\[\Lambda^{2}_{c}=1.99\,\, 10^{-13},\]
\be
m^{2}_{3/2}=1.24\,\, 10^{-32}.
\a{e58}\ee
This gives a value of $m_{3/2}=\,249\, GeV$ for the gravitino
mass and $\Lambda_{c}=\,8.93\,10^{11}\,GeV$ for the condensation
scale both of which are phenomenologically realistic. The
corresponding value for the gauge coupling at the
compactification scale is $g^{2}=\frac{1}{2.65}$.

As discussed above confinement effects may cause the spectrum
determining one loop effects to differ from that used above so
we turn to a consideration of how these effects may alter our
conclusions. Since we do not know the details of this spectrum
we
will assume that, as an average of these effects, the states will
get a common
(supersymmetric) mass $m_{con}$ which
we expect to be proportional to the condensation scale ($
m_{con}=a\,\Lambda_{c}$). We can now minimize the scalar
potential
including these confinement effects. At the minimum the value
of the modulus $S$ and the gravitino mass are:
\[
S_{r}\simeq\,8\pi\sqrt{\frac{1}{ng}}\frac{1}{\sqrt{1-a\,ln(\frac
{1+a}{a})}}
\]
\be
m'^{2}_{3/2}= m^{2}_{3/2}\,(1-a\,ln(\frac{1+a}{a}))^{3/2}
\la{p11}\ee
with $m^{2}_{3/2}$ given in eq.({\ref{e57}). We see
from eq.(\ref{p11}) that the effects of the confinement mass is to
shift (increase) the v.e.v. of $S$. Clearly it is still possible to
obtain phenomenologically interesting results for reasonable values of
$a$ (i.e. $a \sim O(1)$).

In all the solutions so far obtained the value of the potential
at the minimum is negative corresponding to a non-vanishing
cosmological
constant of order $\Lambda_{c}^{4}$, many orders of magnitude
larger than present bounds. This problem is shared by all
attempts to generate supersymmetry breaking and we have nothing
new to add to the discussion.  It is possible to cancel the
cosmological constant by modifying the superpotential and fine
tuning  but we postpone a discussion of this  and of the
implications of the supersymmetry breaking discussed here for the
moduli and scalar fields to a subsequent paper \cite{23}.

\section{Conclusions and Summary}

If gaugino condensate forms in a N=1 supergravity theory
supersymmetry will be broken. The condition that the Goldstone
mode, associated with the spontaneous breaking of the R symmetry
by the gaugino condensate, be coupled in an N=1 locally
supersymmetric way leads to a highly restricted form of the
effective Lagrangian describing this mode. We have determined
this Lagrangian and demonstrated that it leads to a description
of the effective potential describing the gaugino condensate
equivalent to the ``truncated'' and  ``effective'' superpotential
approaches.

 However, the description in terms of the Goldstone mode shows
that the effective potential should be corrected by radiative
effects which may present approach allows for the study of the
gap equation and it was found that a
 strong effective coupling between the gauginos is dynamically
preferred and  a gaugino condensate
 energetically favoured. After imposing a cut-off on the
effective coupling consistent with dimensional analysis we
determine the fine structure constant at the compactification
scale as  well as the magnitude of the hierarchy between the
gravitino and the Planck masses. They are largely determined by
the $\beta$-function and dimension of the
 hidden sector gauge group. Different phenomenologically
interesting solutions  are possible and we showed, as an example,
how a choice based on a SU(5) group gave very reasonable values
both for the gauge coupling and the mass hierarchy.

The work of A.M. has been supported by U.N.A.M.
(Mexico) and ORS (U.K.).

\end{document}